# Effect of Electric Field on the Band Structure of Graphene/BN and BN/BN Bilayers

Radhakrishnan Balu[1*], Xiaoliang Zhong[2], Ravindra Pandey[2], and Shashi P. Karna[1]

[1]US Army Research Laboratory, Weapons and Materials Research Directorate, ATTN: RDRL-WM, Aberdeen Proving Ground, Maryland 21005-5069, USA

[2]Department of Physics, Michigan Technological University, Houghton, Michigan, 49931, USA

(August 8, 2011)

[*]Corresponding author:

Radhakrishnan Balu (Radhakrishnan.balu.ctr@mail.mil)

## Abstract

Effect of electric field on the band structures of graphene/BN and BN/BN bilayers is investigated within the framework of density functional theory. A relatively large degree of modulation is predicted for the graphene/BN bilayer. The calculated band gap of the graphene/BN bilayer increases with applied electric field, which is not the case of BN/BN bilayer; its band gap decreases with the applied field. Both features in the bilayers considered appear to be related to the nature of the conduction band and the redistribution of the electronic charge of the bilayer systems under the influence of electric field

## 1.0 Introduction

Periodic systems in two-dimensional arrangements have received a great deal of attention due to their novel electronic, electrical, and mechanical properties [1-4]. For example, graphene, a planar structure of hexagonal carbon rings shows extraordinary electrical properties in its pristine form and has been the subject of numerous recent studies[5]. Its band structure shows zero gap at the Dirac point, which can be opened by external perturbations such as the application of strain, electric field and chemical modification[6,7]. The band gaps of both bilayer (BLG) and trilayer (TLG) configurations of graphene are tunable with an applied electric field[8] and are dependent on the symmetry. For example, in the case of TLG even though inversion symmetry is broken by applied electric field hexagonal stacking has no band gap, bernal stacking has a tiny gap and only the rhombohedral stacking has sizeable gap. Structurally similar to graphene, hexagonal BN (h-BN) has also received considerable attention as a potential material for nano-scale electronics applications[9] due to its enhanced chemical, thermal, and mechanical stabilities. The enhanced stability of BN results from the presence of sigma bonded covalent bonds.

In the case of bilayer graphene, recent studies [10,11] have shown that band gap increases from zero to 230 meV at 3 V/nm. Such a possibility opens the door for bilayer graphene in switchable electronic devices. Since BN is chemically more stable and already has a sizeable band gap, it is of interest to investigate the effect of electric field on the graphene/BN and BN/BN band gaps, both for advancing fundamental understanding of the electronic structures of these important nano-scale materials and also for their potential applications in switchable devices. In this Letter, we report the results of such a study of the effect of electric field on the band structures of graphene/BN and BN/BN bilayers investigated by first-principles density functional theory approach. Our results suggest both graphene/BN and BN/BN layers exhibit modulation of the band structure by electric field. However, significant qualitative and quantitative differences are noted.

## 2.0 Computational Method

Calculations were performed using the full-potential linearized augmented plane wave (FLAPW) method within the framework of local density approximation (LDA) - density functional theory (DFT). In the FLAPW method, the crystal region is split between non overlapping muffin-tin spheres around nuclei and interstitial regions. The plane wave basis set is used to describe the interstitial region and radial functions in the muffin-tins to account for the sharply changing potential near the nuclei. The linearized augmented plane wave-based methods are known to give accurate electronic structure description of solids[12]. The LDA-DFT has been previously used to obtain reliable results for graphitic and h-BN band structures [13-15].

Two dimensional slab geometries were used for the systems studied; the interlayer spacing in the AB (Boron) stacking for graphene/BN and BN/BN equilibrium configurations (Figure 1) are 3.022 Å and 3.071 Å, respectively and are taken from the work of Zhong *et al*[9]. The AB stacking for graphene/BN systems was chosen as it is the energetically favored configuration over AA stacking at the LDA-DFT[9] level of theory. The electric field was applied in the direction perpendicular to the planes of the bilayer by setting up two plates of opposite charges on either side of the bilayer systems as implemented in FLEUR[16] electronic structure code. The charge-density plots were generated using XcrysDen[17] visualization package.

**3.0 Results and Discussion**

Figure 2 shows the band structures of graphene/BN and BN/BN bilayer systems. As expected, the graphene/BN and BN/BN bilayer band structures exhibit sharp distinctions with each other. The zero-field band structure for the graphene/BN bilayer is dominated by the bands associated with the carbon atoms near the Fermi level, with a very small gap of ~ 104 meV at the Dirac point. The bands near the Fermi level exhibit linear dispersion characteristic of graphene with their slopes less steep than bilayer graphene due to the presence of BN layer. The presence of non-zero gap implies lifting of degeneracy of the bands at the K point. In contrast, the BN/BN bilayer exhibits a nearly two-orders of higher direct gap of ~ 4.6 eV at the Dirac point. This is in accordance with earlier calculations[9].

The change in the band gaps of graphene/BN and BN/BN bilayers is shown in Figure 3. It is clear from this figure that both bilayer systems exhibit modulation of their band gaps by the external field. While the graphene/BN bilayer band gap shows an increase with increasing external field, opposite is the case with the BN/BN bilayer. Further, the relative change in the graphene/BN bandgap appears to be much larger than that in the BN/BN bilayer, although the magnitude of the change in the latter is much higher than the former. The decrease in the gap with increase in the electric field in the BN/BN bilayer is consistent with the previously reported results on nanotubes of BN[18]. The magnitude of the band gap with the electric field appears to vary in a linear way, though a quadratic variation appears for small electric field (<1.2 V/nm) for the BN/BN bilayer system.

An examination of the band structure reveals that the conduction bands are affected more strongly than the valence bands (Table I) by the external electric field for both the graphene/BN and BN/BN bilayers. For the graphene/BN system, the valence and conductions bands are pushed higher with increasing field with a net increase in the gap. In contrast, the conduction band of BN/BN bilayer is pushed towards the Fermi level with increasing field. While the valence band is pushed higher with increased field until E=2.0 V/nm, decreased at V=2.5 V/nm and then monotonically increased from V=2.5 V/nm onwards. This results in an overall decrease in the band gap. The effect of applied field on bands closer to Fermi level can be summarized as moving in the same direction for graphene/BN bilayer, moving in opposite direction that is both bands move towards the Fermi level, for the BN/BN bilayer but in all cases the change is linear with respect to the applied field. This is similar to the observation on trilayer graphene under the influence of electric field[7] where all the bands closer to Fermi level change linearly but the direction of change depends on the symmetry of the system.

In order to understand the effect of the applied electric field we plotted the difference charge density *(*between E=0 V/nm and E=2.5 V/nm) associated with valence and conduction bands near Fermi level ($E_F$ -2 eV < $E_F$ < $E_F$ +2 eV) in Figure 4. Here, we chose a top view of the bilayer system with graphene being the top layer of graphene/BN system. The contours are in the linear scale with blue regions correspond to decrease in charge density and red regions indicate increase in charge density from zero bias to E=2.5 V/nm. The plane of the contours is chosen exactly mid way between the two layers. For the valence band there is an increase in charge density at the nitrogen atoms and decrease in charge density at the boron atoms. The reverse is true in the case of conduction band. Figure 5 shows the same plot for BN/BN bilayer system *(*between E=0 V/nm and E=2.5 V/nm) associated with valence and conduction bands near Fermi level ($E_F$ -2 eV < $E_F$ < $E_F$ +5 eV). No significant change in the charge density was seen for nitrogen atoms, though there is a slight change seen for the boron atomic centers.

## 4.0 Summary

We have investigated the effect of the electric field on the graphene/BN and BN/BN bilayer systems. The calculated results show the band gap in graphene/BN bilayer is dominated by graphene features and is relatively more tunable than that of the BN/BN bilayer. The calculated results show the band gaps of both, graphene/BN and BN/BN, bilayers to be tunable: +14% in graphene/BN and -5% in BN/BN bilayer for an applied electric field of 4.0 V/nm. The degree of modulation of the band gap is related to the nature of conduction band together with redistribution of the electronic charge of the bilayer systems under the influence of electric field.

**Acknowledgements:**

Calculations were performed using the DOD Supercomputing Resource Centers (DSRCs) located at the U.S. Army Engineer Research and Development Center. The work at Michigan Technological University was performed under support by the U.S. Army Research Laboratory through Contract No. W911NF-09-2-0026-133417.

Figure Caption:

*Figure 1. A schematic diagram of graphene/BN (a) and BN/BN (b) bilayers. The cyan, green, and blue represent carbon, boron and nitrogen atoms, respectively.*

*Figure 2. Band structure of (a) graphene/BN and (b) BN/BN bilayers at zero bias. Zero of the energy is set to the valence band maximum. The t in (2a) shows the band gap near the K point of graphene/BN bilayer.*

*Figure 3. Electric-field induced variation of the band gap of graphene/BN and BN/BN bilayers.*

*Figure 4. Differnce charge distribution (between E=0 and E≠0) of (a) valence band and (b) conduction band of Graphene/BN bilayer.*

*Figure 5. Differnce charge distribution (between E=0 and E≠0) of (a) valence band and (b) conduction band of BN/BN bilayer.*

*Table 1. Energies of bands at Dirac point of the graphene/BN and BN/BN bilayers.*

*Figure 1. Balu et al.*

(a) 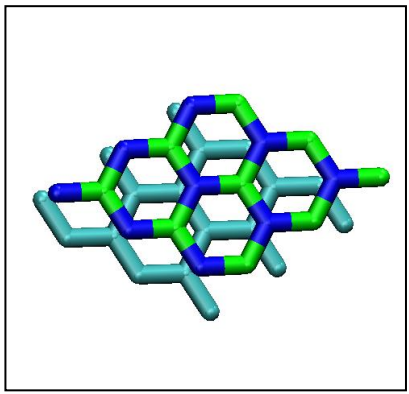    (b) 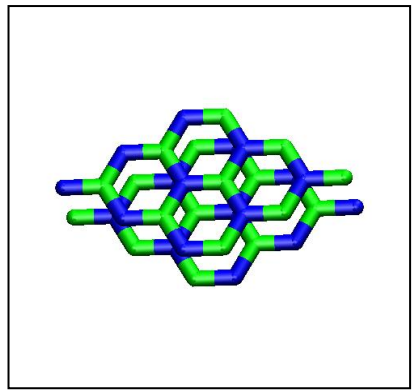

Figure 2: Balu et al.

(a) (b)

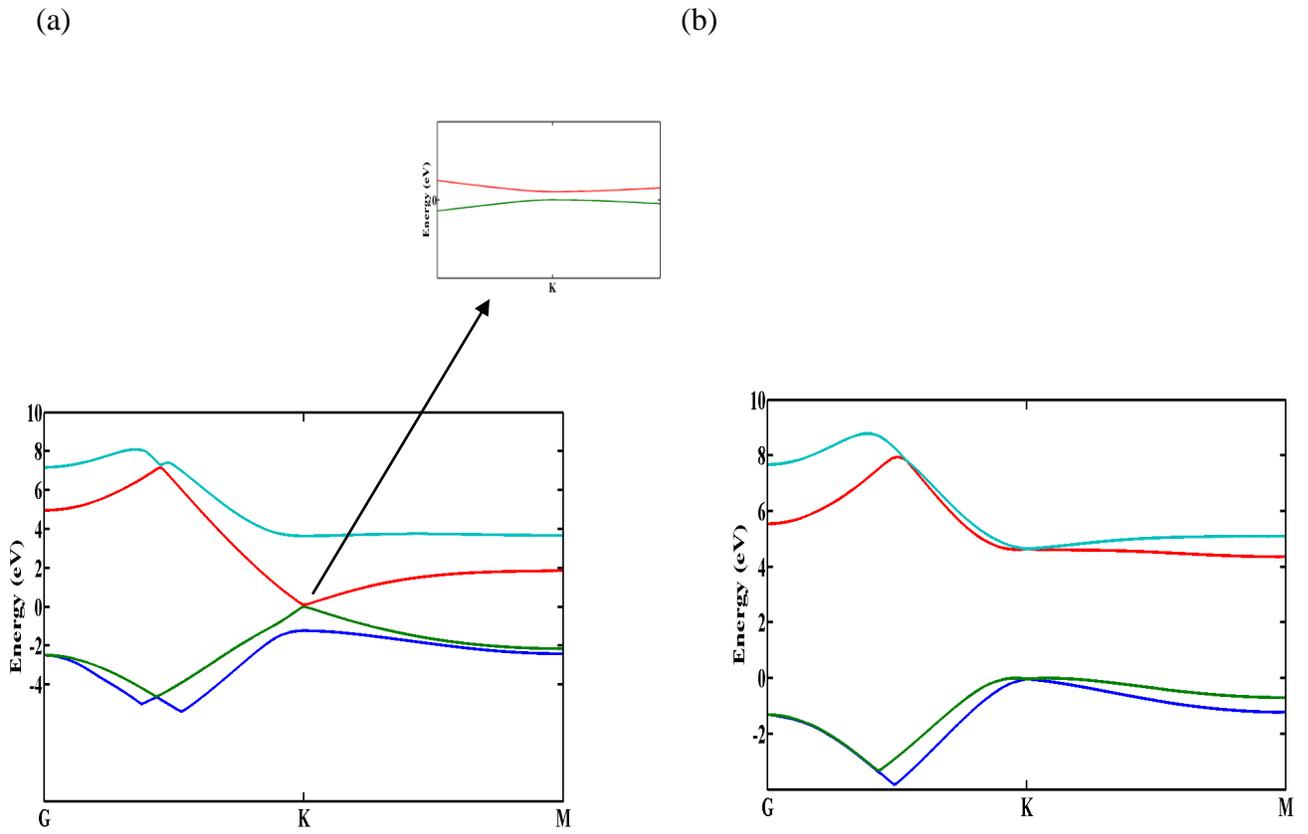

*Figure 3. Balu et al.*

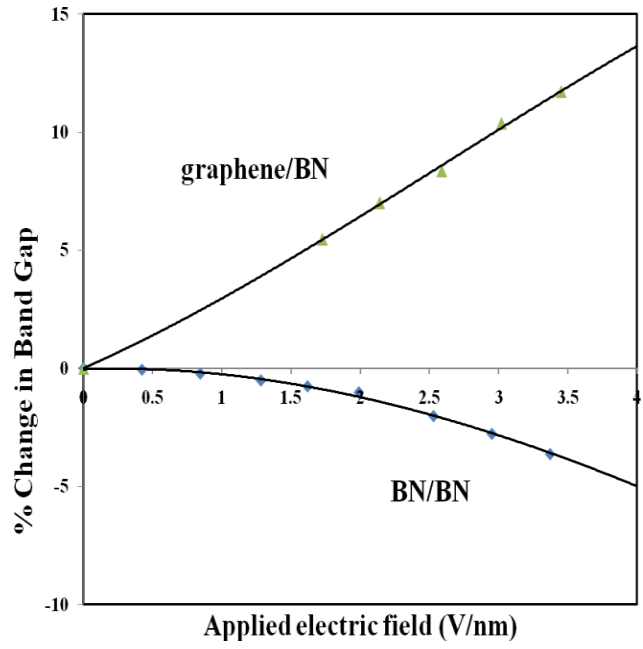

(a) 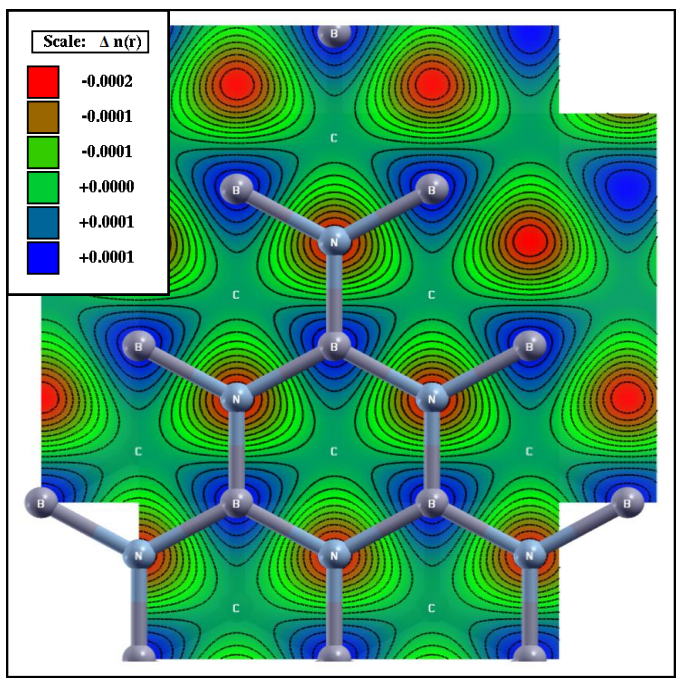 (b) 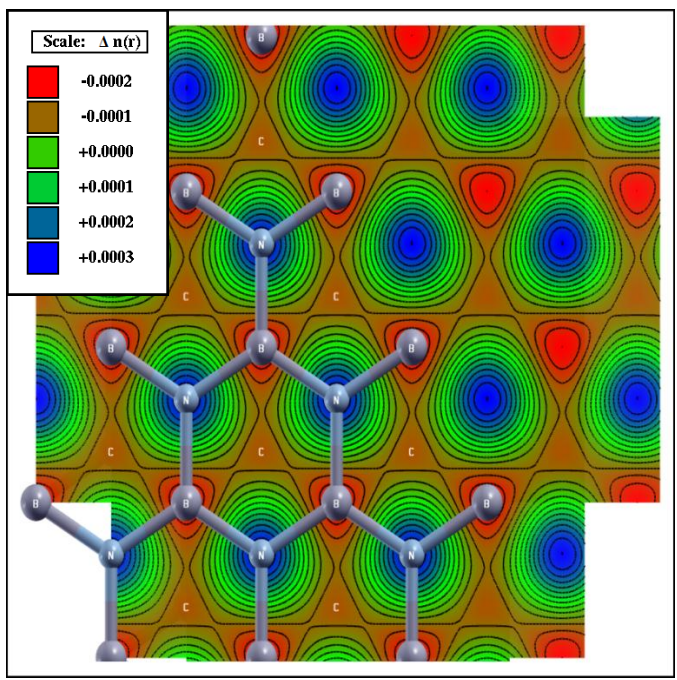
2

*Figure 5. Balu et al.*

(a) 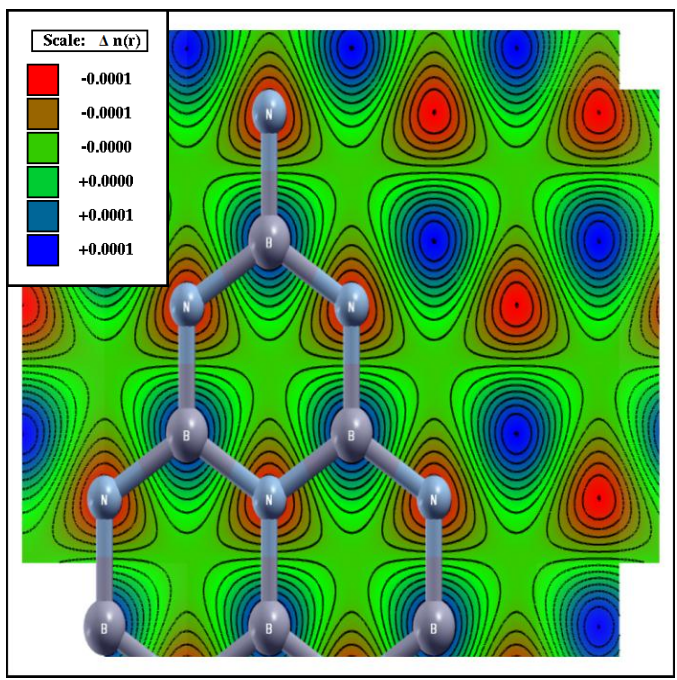

(b) 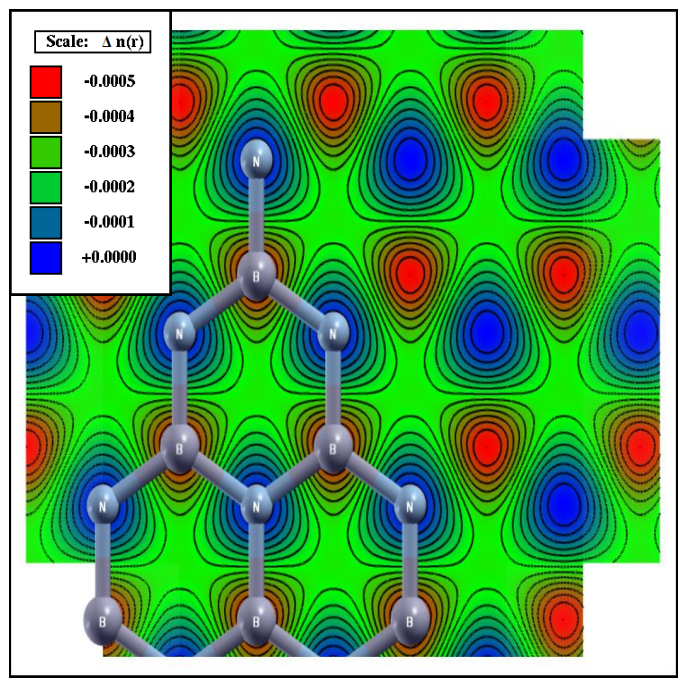

*Table I. Energies of bands at Dirac point of the graphene/BN and BN/BN bilayers.*

| SYSTEM | EFIELD (V/nm) | VB (eV) | CB (eV) | Band gap (eV) |
|---|---|---|---|---|
| graphene/BN | 0.0 | 2.157 | 2.262 | 0.104 |
| graphene/BN | 1.7 | 2.219 | 2.329 | 0.109 |
| graphene/BN | 2.1 | 2.227 | 2.339 | 0.111 |
| graphene/BN | 2.5 | 2.235 | 2.349 | 0.112 |
| graphene/BN | 3.0 | 2.241 | 2.357 | 0.115 |
| graphene/BN | 3.4 | 2.245 | 2.362 | 0.116 |
| graphene/BN | 4.1 | 2.251 | 2.370 | 0.118 |
| BN/BN | 0.0 | 0.580 | 5.174 | 4.594 |
| BN/BN | 0.4 | 0.581 | 5.172 | 4.591 |
| BN/BN | 0.8 | 0.585 | 5.169 | 4.584 |
| BN/BN | 1.2 | 0.592 | 5.162 | 4.570 |
| BN/BN | 1.6 | 0.602 | 5.159 | 4.557 |
| BN/BN | 2.0 | 0.805 | 5.141 | 4.537 |
| BN/BN | 2.5 | 0.624 | 5.125 | 4.501 |
| BN/BN | 2.9 | 0.636 | 5.102 | 4.466 |
| BN/BN | 3.3 | 0.648 | 5.074 | 4.427 |
| BN/BN | 4.0 | 0.666 | 5.028 | 4.362 |